\documentclass[twocolumn,epsfig]{revtex4}
\usepackage{epsfig,color}
\begin{document}
\title{In-medium effects on $K^{+}$ and $K^{-}$ spectra in lighter systems}

\author{Aman D. Sood$^1$ }
\email{amandsood@gmail.com}
\author{Ch. Hartnack$^1$ }
\author{J\"org Aichelin$^1$ }
\address{
$^1$SUBATECH,
Laboratoire de Physique Subatomique et des
Technologies Associ\'ees \\University of Nantes - IN2P3/CNRS - Ecole des Mines
de Nantes 
4 rue Alfred Kastler, F-44072 Nantes, Cedex 03, France}
\date{\today}

\maketitle

\section*{Introduction}
More than a quarter of a century ago, the first strange particles have been observed in heavy-ion collisions (HIC) at relativistic energies and since then the strangeness production has become a major research field in this domain. HIC are of special interest as strange particles can be produced below the corresponding threshold in elementary nucleon-nucleon collisions. The production of strange mesons ($K^{+}$ and $K^{-}$) has created a lot of interest because it has been proposed to use them to test the properties of the nuclear environment, especially of the nuclear equation of state \cite{Aichelin:1986ss}. One of the key question in the analysis of sub-threshold kaon production is how to obtain information on the properties of strange mesons in dense nuclear matter since simulations show that several observables are sensitive to the in medium properties of mesons \cite{hartphysrep}. 
The principal problem for extracting  precise information on these properties  is, however,  
 that almost all observables depend simultaneously not only on the $K^-$ potential but also on several other 
input quantities which are only vaguely known. They include  the lifetime of the $\Delta$ and the modification of $\sigma_{NN \to K^{\pm}X}$ in the medium, the only theoretically known $\sigma_{N\Delta \to K^+N\Lambda} $ cross section and the little known cross sections for the production  of the $K^-$ in secondary interactions $BY \to BB K^-$ or  $\pi Y \to K^- N$ (where $Y \to \Lambda, \Sigma$) which dominate the $K^-$  production \cite{hartphysrep}. This new reaction channels (occurring only in HIC ) link the $K^-$ to $K^+$ production. Thus all the uncertainties related to the production of $K^+$ are inherited by $K^-$. Here we aim to explore the in-medium effects on the transverse momentum ($p_{T}$) spectra of $K^{+}$ and $K^{-}$ in lighter mass system $^{12}C+^{12}C$. Lighter mass has been chosen so that there is little or no effect of other little or unknown quantites mentioned above. 

In order to check and make this sure 
we have separated the $K^-$ into 2 classes (by tracing back $K^-$ to its corresponding anti strange partner $K^+$).\\ (a) $K^-$ coming directly from  reactions like $BB \to BB K^+ K^-$ called direct contribution  and abbreviated in the figure (Dir)\\
(b) $K^-$ coming from $\pi Y$ or $BY\to K^-$ abbreviated by Y. 

Korpa and Lutz \cite{Korpa:2004ae} have calculated the energy of kaon in medium $\omega({\bf{k}}=0,\rho)$ using a selfconsistent
Bethe Salpether equation.  The result of these calculations can be well approximated by
\begin {eqnarray}
\omega({\bf{k}}=0,\rho)& =&m_{\rm K^+}(\rho)
\nonumber\\
&=& m_{\rm K^+}(\rho =0) (1+\alpha_{\rm K^+} \frac{\rho}{\rho_{0}}) 
\end {eqnarray}
\begin{eqnarray}
 m_{\rm K^-}(\rho )=m_{\rm K^-}(\rho =0) (1+ 
\alpha_{\rm K^-} \frac{\rho}{\rho_{0}}) \label{kpmass}
\end{eqnarray}
with $\alpha_{\rm K^+} = 0.07$ and $\alpha_{\rm K^-} = -0.22$. To study the influence of the potential we multiply
$\alpha_{K^+}$ and $\alpha_{K^-}$ by a artificial factor x. 
For the present study we use IQMD model \cite{hart98}. The details of IQMD program on how
strange particles are described in this approach have been extensively described
in ref. \cite{hartphysrep}. 
\section*{Results and Discussion} 
\begin{figure}[!t] \centering
\vskip 0.5cm
\includegraphics[angle=0,width=6cm]{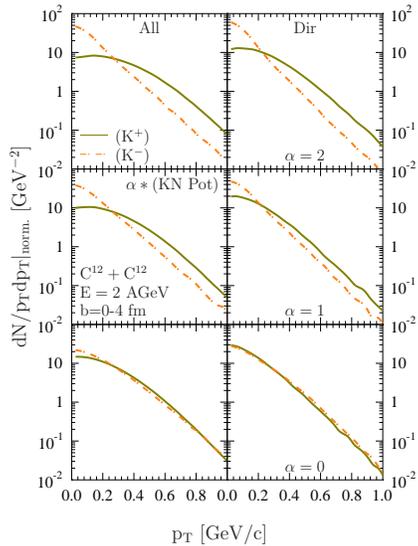}
\caption{\label{fig1} $p_{T}$ spectra of kaons for different 
strengths of KN potential.}
\end{figure}

For the present study we simulate the $^{12}C+^{12}C$ reactions at an incident energy of 2 AGeV 
in the impact parameter range of 0-4 fm. In fig \ref{fig1}, we display the transverse momentum ($p_{T}$) 
spectra of $K^+$ (solid) and $K^-$ (dash dotted line)  for $0\le b\le 4$ fm obtained for CC collisions at $E_{beam}$=2 AGeV.
The K potentials are used for both $K^+$ and $K^-$  and x= 2, 1 and 0 and for the top to the bottom panels.  
The spectra is normalized to 1 so that the $K^+$ and $K^-$ spectra can be directly compared. Left and right panels represent the spectra for all the kaons (labelled as All)  and kaons produced in reaction BB $\to$ BB$K^+$$K^-$ (labelled as Dir), respectively. When we switch off the KN potential (x=0, bottom panel), the shape of the $K^+$ and $K^-$ spectra is almost identical.  We start the discussion with the right panel. For vanishing potential 
spectra is identical for both kaons. 
The fact that also finally the spectra are almost identical indicates that for these light systems rescattering has little influence on the spectral form. When we switch on  the potential we find $ \omega({\bf{k}}=0,\rho) < 500 MeV$ for the $K^-$ whereas  for the $K^+$  we obtain $ \omega({\bf{k}}=0,\rho) > 500 MeV$. When the kaons leave the nucleus they have to get rid
of the excess mass ($K^+$) or they have to acquire mass ($K^-$).  For the $K^+$ a part of this excess mass is converted into kinetic 
energy whereas for the $K^-$ a part of the kinetic energy is converted into mass. (In IQMD  the total momentum and the total energy of all particles is conserved).  Consequently,  the number of $K^-$ ( $K^+$ ) increases (decreases) in the low momentum region causing the different slopes of the spectra for $K^-$ and $K^+$. This effect increases with 
increasing strength of potential (top panel). 

The left panels show that for vanishing  potential  the $K^-$ produced in a secondary collisions have almost the same 
slope as the directly produced. Increasing the potential we see that the effect which we have observed on the right hand
side survives if we include all $K^-$. This is essential because both classes cannot be discriminated experimentally.

\section*{Acknowledgments}
This work has been supported by a grant from Indo-French Centre for the Promotion of Advanced Research (IFCPAR) under project no 4104-1.


\end{document}